\documentclass[runningheads]{llncs}
\usepackage[T1]{fontenc}
\usepackage{graphicx}
\usepackage{tabularx}
\usepackage{multirow}
\usepackage{inconsolata}
\usepackage{tabularx}
\usepackage{graphicx}
\usepackage{amssymb}
\usepackage{amsmath}

\begin{document}
\title{Token-Level Graphs for Short Text Classification}
%
%
\author{Gregor Donabauer \and Udo Kruschwitz\thanks{version as accepted at ECIR 2025 with author information and Github link added}}
\authorrunning{Donabauer and Kruschwitz}
%
\institute{Information Science, University of Regensburg, Regensburg, Germany \\ \email{\{gregor.donabauer,udo.kruschwitz\}@ur.de}}
\maketitle              
\begin{abstract}

The classification of short texts is a common subtask in Information Retrieval (IR). Recent advances in graph machine learning have led to interest in graph-based approaches for low resource scenarios, showing promise in such settings. However, existing methods face limitations such as not accounting for different meanings of the same words or constraints from transductive approaches. We propose an approach which constructs text graphs entirely based on tokens obtained through pre-trained language models (PLMs). By applying a PLM to tokenize and embed the texts when creating the graph(-nodes), our method captures contextual and semantic information, overcomes vocabulary constraints, and allows for context-dependent word meanings. Our approach also makes classification more efficient with reduced parameters compared to classical PLM fine-tuning, resulting in more robust training with few samples. Experimental results demonstrate how our method consistently achieves higher scores or on-par performance with existing methods, presenting an advancement in graph-based text classification techniques. To support reproducibility of our work we make all implementations publicly available to the community\footnote{\url{https://github.com/doGregor/TokenGraph}}.

\keywords{Pre-trained Language Models \and Text Classification \and GNNs.}
\end{abstract}

\section{Introduction}
\label{sec:intro}

Short text classification is considered an important subtask in many Information Retrieval (IR) pipelines, and examples include assigning a label or category to snippets in search results or to tweets, status updates, comments, and reviews from social platforms \cite{sun2012short}. Over the years, various methods have been proposed to improve the accuracy and efficiency of this task and fine-tuning of pre-trained language models (PLMs) is among the most popular approaches nowadays \cite{devlin_et_al_2019}. However, in situations where labeled data is scarce, such models often perform poorly or unstable \cite{zhang2021revisiting}. Driven by progress in graph machine learning \cite{kipf2017semisupervised,hamilton_et_al_2017}, it has recently become increasingly popular to employ graph-based approaches on the task of short text classification that show more robust performance in low resource settings e.g., \cite{wang_et_al_2021,zheng_et_al_2022,yadati_2024}.

However, many existing approaches are facing certain limitations: Some rely on large networks of text documents to classify nodes within these graphs, making the approach transductive \cite{linmei_et_al_2019,wang_et_al_2021,lin_etal_2021_bertgcn,xiao_etal_2021_bert4gcn}. Others adopt an inductive approach but are constrained by different drawbacks, such as stopword removal, which further truncates already concise short texts \cite{zhang_et_al_2020,zheng_et_al_2022,wu_et_al_2023}. Additionally, most methods depend on word representations (e.g., pre-trained embeddings like GloVe \cite{pennington-etal-2014-glove} and word2vec \cite{word2vec}, or one-hot vectors), lacking the ability to capture context-dependent word meanings \cite{zhang_et_al_2020,ding_et_al_2020}. 

To address these constraints, we introduce an approach wherein text preprocessing for constructing text graphs is based on PLM tokenization. While the idea of combining PLMs and graph-based text classification is not novel \cite{lin_etal_2021_bertgcn,xiao_etal_2021_bert4gcn} our method allows to go beyond limitations of existing approaches (such as transductive settings) by breaking the construction of text graphs down to the token level of PLMs. We propose to feed text samples through the PLM's tokenizer, breaking them into word units. Subsequently, these tokens are linked within a $n$-hop fully connected graph, with each text sample representing a distinct graph. The embeddings of token nodes within each graph are derived by feeding the tokens through the PLM.

As a result, our approach represents an inductive setting (each sample is represented by a distinct graph) and maintains the positive implications of using PLMs for text representation. We can embed nearly any text since the vocabulary of the PLM is more fine-grained (tokens instead of words) which allows us to also tokenize unknown words into smaller word units. Furthermore, the embeddings of these tokens are context-dependent, as they are embedded within the complete sequence using the PLM. The embeddings also represent semantic information between words, inherent in the PLM's weights. During training our approach easily allows to reduce the number of parameters in the graph neural network (GNN) that can be used to classify the token-graphs compared to fine-tuning a PLM. This makes overfitting less probable and improves the training robustness. Nonetheless, the aggregation process remains similar to fine-tuning a PLM, as it has been shown that Transformers operating on token sequences are essentially a variant of GNNs \cite{joshi2020transformers,VELICKOVIC_2023}.

We conduct multiple experiments on diverse datasets and compare our approach to state-of-the-art systems in the field. Our findings demonstrate that we constantly achieve better or competitive performance compared to existing methods in the domain.

\section{Related Work}

\subsection{Graph Neural Networks (GNNs)}

Graphs, or networks, offer a unique way to model any interactions between components in a system \cite[Ch.1]{networks_2013}. This idea goes beyond well-defined inputs, such as sequences (e.g. text) or grid-structures (e.g. images) and thus requires another type of encoder than well-defined neural networks, such as recurrent or convolutional neural networks \cite[Ch.5]{hamilton_networks_2020}. As a solution, graph neural networks that can operate on graph-structured data have been proposed \cite[Ch.5]{hamilton_networks_2020}. Many variations of GNNs have been introduced over the last years, this includes Graph Convolutional Networks \cite{defferrard_et_al_2016,kipf2017semisupervised}, GraphSAGE \cite{hamilton_et_al_2017} and Graph Attention Networks \cite{velickovic2018graph}. Due to their unique abilities, they have also sparked interest in the NLP/IR community where a variety of problems can also benefit from being modelled as a graph structure \cite{wu_et_al_2021,wu_et_al_2023,donabauer2023exploring}. This also includes text classification \cite{wu_et_al_2023}.

\subsection{GNNs for Short Text Classification}

Several recent approaches have explored graph-based techniques for short text classification. Linmei et al. \cite{linmei_et_al_2019} use heterogeneous graph attention networks to aggregate information over a heterogeneous network where nodes represent text samples, entities and topics. Similarly, Wang et al. \cite{wang_et_al_2021} use heterogeneous networks comprising word-level graphs (including POS-tag and entity graphs) as well as a document-level graph for short text similarity to capture similarities between short texts in the corpus-level graph. Ding et al. \cite{ding_et_al_2020} represent documents as hypergraphs where nodes represent words and different hyperedge types account for co-occurrences of words and topic-related high-order correlations. Another approach based on homogeneous word graphs also modelling co-occurrences of words as edges within a fixed-size sliding window is presented by Zhang et al. \cite{zhang_et_al_2020}. To inject more corpora-level knowledge into graph-based text classification, Zheng et al. \cite{zheng_et_al_2022} construct a word co-occurrence graph of common words based on such a corpus. Text samples can then be modelled as the average of these embeddings of words appearing in the corpus graph and then be classified. Based on that approach, Yatadi \cite{yadati_2024} introduces a refined GNN architecture that calculates a node-specific radius during aggregation going beyond fixed-hop neighborhoods. Finally, Lin et al. \cite{lin_etal_2021_bertgcn} first fine-tune BERT and then use the model to encode document nodes in a heterogeneous graphs of words and documents. Due to the transductive setup, the GCN used to aggregate the information can also observe test nodes during training.

In conclusion, research in graph machine learning has recently pushed graph-based short text classification and resulted in various new approaches. However, these methods are often transductive or have limitations in representing the text completely. 
We want to address these issues by breaking down preprocessing to sample-wise graphs and the token-level of PLMs as this results in several positive implications as described in Section \ref{sec:intro}.

\section{Methodology and Experiments}

\subsection{Token-Based Graph Construction}
\label{sec:tokengraph}

In general, each text sample within the dataset is transformed into a graph. The number of nodes within each graph corresponds to the number of tokens obtained by the PLM's tokenizer during the text tokenization process. The features of each node are derived from feeding the full token sequence through the PLM. All nodes are connected within their $n$-hop neighborhood. We describe this procedure in more detail below.

\textbf{Text Preprocessing:} Let $T$ be a text sample in our corpus. We denote the tokenization function (using a PLM's tokenizer) as $\phi_{Tokenizer}$. This function takes a text sample $T$ and produces a sequence of tokens: $\phi_{Tokenizer}(T)=[t_{1},t_{2},...,t_{n}]$ that we call $S_{T}$ and where $t_{i}$ represents the $i$th token in the sequence. The specific process of tokenization depends on the PLM's tokenizer but in general it includes breaking the input into words or more fine-grained word units \cite{song-etal-2021-fast}. The tokenizer is also able to handle special tokens like punctutaion marks, special characters, URLs, etc. which are removed in most existing methods due to preprocessing steps such as stopword removal (e.g., \cite{zhang_et_al_2020,zheng_et_al_2022}).

\textbf{Graph Construction:} After tokenization we can transform the token sequence $S_{T}$ into a graph. Each graph can be formulated as $G = (V,E)$ where $|V|=|S_{T}|$. Each node in the graph is associated with a $d$-dimensional vector $x_{v} \in \mathbb{R}^{d}$. The exact representations are obtained by embedding the tokenized sequence by feeding it in the language model $X = PLM(S_{T})$. Note that this process results in contextual embeddings where tokens associated with the same ID in the vocabulary have different representations depending on the words/tokens surrounding them. Each edge in our graph connects nodes within a $n$-hop neighborhood in the sequence. We iterate over each token $t_{i}$ and connect it to other tokens within a $n$-hop neighborhood, e.g. for $n=2$, $t_{i-2},t_{i-1},t_{i+1},t_{i+2}$, which results in $E=\{(u,v)|u,v \in V, \text{ and } |u-v| \leq 2\}$. We found $n=1$ to work best\footnote{We ran various ablation studies and use the settings with best performance across all datasets, i.e. $n$-hop, GNN type, layer dimension. For details see our Github.}.

\subsection{Experimental Setup}

\textbf{Data:} We evaluate our approach on multiple short text classification datasets that have also been used in previous research on graph-based short text classification (e.g. \cite{zheng_et_al_2022,yadati_2024}) and are considered benchmarks in the domain \cite{wu_et_al_2023}. In detail the datasets we used in our experiments include \textit{Twitter}\footnote{\url{https://www.nltk.org/howto/twitter.html\#corpus\_reader}} (sentiment classification of tweets), \textit{MR} \cite{pang_and_lee_2005} (sentiment classification of movie reviews), \textit{Snippets} \cite{phan_et_al_2008} (classifying Google search snippets by the topic of search) and finally \textit{TagMyNews} \cite{vitale_et_al_2012} (classifying titles from RSS feeds by topic). More details on the statistics of the datasets can be found in our Github repository.

\textbf{PLMs:} For tokenization and embedding we use BERT \cite{devlin_et_al_2019} in its base and uncased version for the \textit{MR}, \textit{Snippets} and  \textit{TagMyNews} datasets as well as a variant that was pre-trained on Twitter-specific data \cite{zhang_et_al_2023} for the \textit{Twitter} dataset. As both models have a hidden dimension of 768 this is also the dimension of the token embeddings representing the initial graph node features.

\textbf{GNN Architecture:} We use a two-layer GNN to aggregate the text-graph representations. We set the dimension per layers to 128 (compared to 768 as the initial dimension) which helps reducing the embedding size during graph convolution and thus the number of model parameters. In detail, we use Graph Attention Convolution layers \cite{velickovic2018graph} with a single attention head. Afterwards we perform mean pooling over all nodes in each graph and feed the representation through a linear layer. Implementations are based on PyTorch Geometric \cite{fey2019fast}.

\begin{table*}[ht!]
\centering
\begin{tabularx}{\textwidth}{|X|ll|ll|ll|ll|}
\hline
\textbf{Dataset $\rightarrow$} & \multicolumn{2}{c|}{\textbf{Twitter}} & \multicolumn{2}{c|}{\textbf{MR}} & \multicolumn{2}{c|}{\textbf{Snippets}} & \multicolumn{2}{c|}{\textbf{TagMyNews}}\\
\hline
\textbf{Model $\downarrow$} & Acc. & F1 & Acc. & F1 & Acc. & F1 & Acc. & F1\\
\hline
TFIDF+SVM & 0.578 & 0.565 & 0.547 & 0.541 & 0.642 & 0.638 & 0.342 & 0.328 \\
LDA+SVM & 0.527 & 0.491 & 0.519 & 0.510 & 0.302 & 0.287 & 0.215 & 0.182 \\
WideMLP & 0.576 & 0.565 & 0.531 & 0.514 & 0.496 & 0.487 & 0.248 & 0.240 \\
HGAT & 0.549 & 0.525 & 0.522 & 0.485 & 0.626 & 0.620 & OOM & OOM \\
HyperGAT & 0.561 & 0.499 & 0.516 & 0.448 & 0.349 & 0.348 & 0.244 & 0.178 \\
TextING & 0.618 & 0.608 & 0.587 & 0.583 & 0.763 & 0.757 & 0.608 & 0.572 \\
SGR & 0.625 & 0.625 & 0.627 & 0.627 & 0.812 & 0.811 & 0.675 & 0.637 \\
NGR & 0.624 & 0.628 & 0.626 & 0.629 & 0.814 & 0.819 & 0.675 & 0.639 \\
\hline
Fine-Tuned PLM & 0.780 & 0.780 & 0.565* & 0.536* & 0.645* & 0.639* & 0.457* & 0.418* \\
SimpleSTC & 0.484* & 0.477* & 0.522* & 0.517* & 0.693 & 0.650 & 0.528 & 0.478 \\
SHINE & 0.660 & 0.657 & 0.640 & 0.629 & 0.789 & 0.785 & 0.554 & 0.544 \\
BertGCN & 0.534* & 0.517* & 0.666 & 0.662 & \textbf{0.871} & \textbf{0.864} & \textbf{0.765} & \textbf{0.729} \\
\hline
Token-Based Graph & \textbf{0.837} & \textbf{0.837} & \textbf{0.702} & \textbf{0.702} & 0.804 & 0.797 & 0.699 & 0.668 \\
\hline
\end{tabularx}
\caption{\label{table:results}
Baseline and token-based graph results on all datasets. Significant differences compared to token-based graphs using $t$-tests and Bonferroni correction at $p < 0.05$ are marked with *. The results in the top part of the Table are reported from the original paper and have therefore not been tested for statistical significance.
}
\end{table*}

\textbf{Training:} In line with previous work, we use exactly 20 samples per class for both model training and validation and all remaining samples for testing (e.g. compare \cite{zheng_et_al_2022,yadati_2024}). To ensure robustness, we repeat this setup five times and report on the average results achieved on the test set. As each sample in the dataset can be interpreted as an individual graph, we use a batch-size of 8 graphs which are shuffled during training. Training is executed in 200 epochs with a learning-rate of $0.00005$. After each epoch we evaluate the model's performance on the validation set and select the best-performing one for evaluation on the test set at the end of training. All token-graph experiments were executed without a GPU. Baselines were executed with one Nvidia RTX A6000 (48GB VRAM). This further underscores the lightweight computational costs of our method.

\textbf{Baselines:} We compare our approach to strong, recent baselines that also propose graph-based short text classification approaches. Specifically, this includes SHINE\footnote{\url{https://github.com/tata1661/SHINE-EMNLP21}} \cite{wang_et_al_2021}, SimpleSTC\footnote{\url{https://github.com/tata1661/SimpleSTC-EMNLP22/}} \cite{zheng_et_al_2022}, and BertGCN\footnote{\url{https://github.com/ZeroRin/BertGCN}} \cite{lin_etal_2021_bertgcn}. We also wanted to rerun SGR and NGR \cite{yadati_2024} but could not get hold of the implementations which is why we will simply report the scores from the paper. For better comparability, we additionally fine-tune the PLMs we utilized for tokenization and node feature extraction. We fine-tune for 3 epochs (compare \cite{devlin_et_al_2019}), utilizing the same setup of 20 samples per class. This approach allows directly comparing PLM fine-tuning with first transforming text samples into graphs using our method. We also report on other baselines that have published results on the same datasets with the same setups as in our study. However, we did not rerun these systems as the results are generally lower than the strong baselines we did rerun for statistical comparison. These baselines include three traditional methods (TFIDF+SVM, LDA+SVM, WideMLP) as well as graph-based approaches HGAT \cite{linmei_et_al_2019}, HyperGAT \cite{ding_et_al_2020} and TextING \cite{zhang_et_al_2020}.

\section{Results}

In line with common practice in NLP/IR, we report on accuracy and macro F1 scores for all setups \cite[Ch. 4]{Jurafsky24Speech}. All values are the average results obtained by running the setup 5 times with the same training/validation/test sets which is identical to the setup reported in the baselines.

While on the \textit{Snippets} and \textit{TagMyNews} datasets BertGCN (transductive approach) performs best, our token-based inductive method achieves the highest scores on the \textit{Twitter} and \textit{MR} datasets. \textit{Snippets} (8 classes) and \textit{TagMyNews} (7 classes) are the datasets where the most samples are used for training (20 samples per class). On data with only two classes and thus fewer training samples (\textit{Twitter} and \textit{MR}) the token-based approach performs notably better. For comparison, PLM fine-tuning only results in good performance on the \textit{Twitter} dataset. On all other datasets we observe low performance when fine-tuning BERT.

\section{Discussion and Conclusion}

As mentioned, PLM fine-tuning in our experiments only achieves good performance on the \textit{Twitter} dataset.
This is likely because the PLM we used for this dataset was pre-trained on domain-specific data, while BERT, used for the other datasets, was pre-trained on more general text corpora.
The advantage of domain-specific pre-training is also exploited by our PLM-based graph construction method, which results in even higher scores.


As mentioned, BertGCN shows better performance compared to our token-level graphs on datasets where a overall higher number of training samples is used. However, we observe our method to perform much more stable across all datasets, also such with minimal number of training samples. In general, our token-based graph setup also does not observe test samples as context in the full graph during training (inductive vs. transductive).

Additionally, our findings suggest that our method achieves better results on datasets where more context is available in each text sample, such as \textit{Twitter} and \textit{MR}. Instead, samples in the \textit{Snippets} dataset often can be interpreted as a list of isolated words lacking substantial contextual meaning e.g., \textit{``ics raleigh ibm ibm internet connection software''}.

Our easy-to-use and lightweight short text classification method combines the advantages of PLMs in representing text data and graph-based approaches that show robust performance in low-resource settings. Overall, our experiments demonstrate consistent improvements or competitive performance compared to strong, recent baselines.

\bibliographystyle{splncs04}
\bibliography{references}

\end{document}